\documentclass[aps,pre,reprint,showkeys]{revtex4-1}
\usepackage[T1]{fontenc}
\usepackage[latin9]{inputenc}
\usepackage{pifont}
\usepackage[amssymb,cdot]{SIunits}
\usepackage[pdftex,usenames]{color}

\usepackage{graphicx}
\usepackage{amssymb,amsfonts,amsmath,stmaryrd}
\usepackage[amssymb,cdot]{SIunits}

\DeclareMathOperator{\e}{e}

\renewcommand*{\Im}{\operatorname{Im}}
\newcommand{\pdiff}[2]{\frac{\partial#1}{\partial#2}}

\newcommand{\dash}[1]{{#1}^\prime}

\begin{document}

\title{Nonlinear dynamics of the mammalian inner ear}

\author{Robert Szalai}
\email{r.szalai@bristol.ac.uk}
\author{Alan R. Champneys}
\author{Martin Homer}
\affiliation{University of Bristol, Dept.~Engineering Mathematics, Merchant Venturers Building, Woodland Road, Bristol BS8 1UB, UK}

\begin{abstract}
A simple nonlinear transmission-line model of the cochlea with
longitudinal coupling is introduced that can reproduce Basilar
membrane response and neural tuning in the chinchilla.  It is found
that the middle ear has little effect on cochlear resonances, and
hence conclude that the theory of coherent reflections is not
applicable to the model. The model also provides an explanation of the
emergence of spontaneous otoacoustic emissions (SOAEs). It is argued
that SOAEs arise from Hopf bifurcations of the transmission-line model
and not from localized instabilities. The paper shows that emissions
can become chaotic, intermittent and fragile to perturbations.

\end{abstract}


%\abbreviations{BM Basilar membrane; CF, characteristic frequency;
%FRF, frequency response function;
%OAE, otoacoustic emission;
%OHC outer hair cell; RoG rate of growth;
%OOC, organ of Corti;
%SOAE, spontaneous otoacoustic emission; SPL,
%  sound pressure level}

\maketitle

\noindent\textbf{Significance:} The cochlea is a remarkable device
that out-performs any human-made system; it is sensitive to sounds
over a million-fold intensity and a ten-octave frequency range, and
can distinguish signals separated by microseconds at frequencies only
0.2\% apart. Here we study the mechanisms that make this work. We
present a nonlinear mathematical model that combines the key
physiological processes, including both longitudinal coupling and hair
cell motility, which produces response patterns that agree with
experiments in different animals.  A dynamical systems analysis of the
model allows us challenge existing theories on the source of
spontaneous otoacoustic emissions, suggesting that the entire organ,
rather than localized instabilities, are key.

\vspace{5mm}

The mammalian hearing organ is a sensitive sensory device
that operates at the extremes of physical limits. It is capable of
resolving sound pressure levels just above atmospheric thermal
noise and of discriminating frequencies 0.2\% apart
\cite{Hudspeth2000}. In order to achieve these features the inner ear
employs an active feedback mechanism \cite{Gold1948}. Like any feedback
loop, it is possible for the one in the inner ear to become
unstable. In this paper we show how such an instability can lead to
self-excited oscillations that are emitted from the ear as sound, which we
propose as a mechanism for the generation of spontaneous otoacoustic
emissions (SOAEs) \cite{Kemp1979}. We derive this from a new
mathematical model of the mammalian ear, which includes both active
somatic motility and longitudinal coupling in the cochlea, linked to
the atmosphere via the middle ear.

One widespread explanation of emission generation is due to Shera
\cite{Shera2003}, Shera and Zweig \cite{ZweigShera1995} and Talmadge
{\em et al.}\ \cite{Talmadge1998}. They argue that spontaneous emissions are
wave instabilities that arise because of coherent reflection at the
stapes and at the characteristic frequency (CF) position of the
cochlea; a mechanism that gives rise to standing waves that are
stabilized by the cochlear nonlinearities.
However, this theory is based on linear properties of the cochlea and does not explain how standing waves are stabilized.

Here, we apply the theory of dynamical systems to revisit the question
of SOAEs, which leads to an alternative explanation for emission
generation.  We show that stability loss in the model generically
corresponds to a Hopf bifurcation \cite{Kuznetsov2004}, which can lead
to a sustained periodic emission.  This is different from the
so-called `Hopf ear' \cite{Eguiluz2003}, in that it is an emergent macro-level
property of the cochlear model, with all its many interconnected
elements, rather than being localized in any individual micro-scale
component. Indeed, one advantage of our approach is that there is no
need to assume, {\em a priori}, the mechanism of the stability loss;
bifurcation theory determines the onset of instability irrespective of
its source.
We further predict that these bifurcations are ubiquitous,
provided there is fine-scale spatial variation of the
cochlear parameters. This finding provides
a consistent explanation for widespread observation of
SOAEs; in
realistic operating conditions a bifurcation point is never far away.
As one goes beyond a Hopf bifurcation, the resulting small amplitude
limit-cycle vibrations can grow in amplitude, deform, and undergo
further instabilities and bifurcations. One possible physical
manifestation of such complex motions is in the way emissions can
appear, disappear or change their character as the cochlea undergoes
changes due to damage \cite{Veuillet2001} or aging \cite{Kohler1992},
for example.

While linear cochlear models can also predict instability, they cannot
adequately explain what happens when multiple instabilities produce a
multiple independent periodic motions.  In a linear system, multiple
oscillations can coexist without any effect on each other. In nonlinear
systems, such as the model we present here, we expect more complex
phenomena to arise, such as intermittent or chaotic oscillations to
arise, for which there is some experimental evidence. For example,
Burns \cite{Burns2009} describes experimentally observed short-term
property changes to SOAEs, including their seemingly random sudden
appearance and disappearance. Keefe {et al.}\ \cite{Keefe1990} used time
series analysis methods that could distinguish chaotic spontaneous as
well as stimulated emissions.  It is though perhaps not surprising that
there are not more reports of intermittent or chaotic SOAEs.  To
eliminate environmental noise one must use temporal
averaging. However, averaging can also mask short term spectral
variations and chaos.

The rest of this paper is organized as follows. First, we introduce our
mathematical model. Then we show how the model reproduces chinchilla
BM mechanical data and neural tuning. Applying spatial roughness to
the feedback coefficients of the model, we show that the cochlea
develops resonances. As we increase the size of the roughness
the resonances turn into instabilities and we observe
spontaneous oscillations. We investigate these vibrations in
detail and conclude that our model can produce complicated dynamics in
line with experimental observations.

\section{Nonlinear transmission-line model of the inner ear}
We start from a simple transmission-line model of the cochlea inspired
by the work of Zweig \cite{Zweig1991}. Our model, instead of a delayed
feedback, contains a distributed feed-forward mechanism to account for
the organ of Corti (OOC) structure. This type of active mechanism was
introduced by Geisler \cite{Geisler1993} and Steele and Lim
\cite{SteeleLim1999} and successfully applied in many models.
Moreover, parameters of such a model are easily interpreted in terms of
the longitudinal coupling within the OOC \cite{Yoon2011}.
In \cite{SzalaiJoMMS} together with Epp, we showed how
such spatial coupling can be directly compared with
time-delayed feedback, but with better stability properties.

As is common in modeling the mechanics of the cochlea, we assume that
the fluid in the perilymph is incompressible and inviscid, which
yields a wave equation for the pressure difference $p$ between the
scala tympani and vestibuli, and the Basilar membrane (BM)
displacement $x_\mathrm{bm}$,
\begin{equation}
\ddot{x}_\mathrm{bm}(\xi,t)=\frac{\varepsilon^{2}}{m}\frac{\partial^{2}}{\partial\xi^{2}}p(\xi,t) \label{eq:fluid}
\end{equation}
where dot denotes (partial) differentiation with respect to time $t$.
We nondimensionalise the longitudinal distance $\xi$, so that the length of the cochlea becomes unity, $0 \leqslant \xi \leqslant 1$.
The lumped parameter $\varepsilon$ is a function
of the geometry of the cochlear chambers and the density of the perilymph
fluid, and $m$ is the mass surface density of the BM.

Our model of the BM motion, and its amplification by outer hair cells
(OHCs), is a simplified version of that developed by \'O Maol\'eidigh
and J\"ulicher \cite{Daibhid2010}.
This model accounts for the active nonlinearity of hair bundles, coupled
to hair cell elongation, for a detailed derivation see
\cite{SzalaiNonlin}. To that model we add longitudinal coupling,
as described in \cite{SzalaiJoMMS}.
Specifically, the BM
displacement $x_\mathrm{bm}$ and OHC charge $q$, at longitudinal
distance $\xi$ from the stapes, are defined by the equations
\begin{align}
\ddot{x}_\mathrm{bm}(\xi,t) & = \frac{1}{m} p(\xi,t) - 2\zeta(\xi)\omega_{0}(\xi)\dot{x}_\mathrm{bm}(\xi,t) - \omega_{0}^{2}(\xi) x_\mathrm{bm}(\xi,t) \nonumber\\
 &\quad - \frac{f_{q} \omega_{0}^{2}(\xi)}{L(\xi)} \int_{-L(\xi)}^{L(\xi)}  w\left(\frac{h}{L(\xi)}\right) q \left(\xi - h, t\right)\mathrm{d}h ,\label{eq:BM}\\
\dot{q}(\xi,t) & =-\gamma q(\xi,t)+\nu\dot{x}_{bm}(\xi,t)\nonumber\\
&\quad +I_\mathrm{hb}\left(P_\mathrm{O}\left(\Delta\dot{x}_\mathrm{bm}(\xi,t)\right) - P_\mathrm{O}(0)\right).\label{eq:charge}
\end{align}
The BM motion has natural frequency
$\omega_{0}$ and relative damping $\zeta$, both of which we assume
to depend on position $\xi$. The BM is forced
both by the cochlear pressure difference $p$
at position $\xi$, and also by the OHCs which exert a force due to the
unique protein called prestin \cite{Ashmore2008} embedded in its
lateral wall.  We assume that the effect of the OHCs is longitudinally
distributed, due to the push-pull mechanism of the OOC
\cite{Yoon2009,Yoon2011}, giving rise to the integral term on the
right-hand side of \eqref{eq:BM}. The longitudinal convolution kernel
$w(x)$ is assumed to have a strong positive feed-forward and a weak negative
feed-backward component,
\[
w(x) = \e^{-\beta (x-\delta_g)^2} \sin (x - \delta_o),
\]
and a characteristic feed-forward distance is given by
$L(\xi) = L_0 \mathrm{e}^{\mu_L \xi}$, that is about twice as long at
the apex than at the base.  Note that the kernal $w$ resembles
a strongly damped wave and therefore might alternatively represent
a secondary  transmission line generated in the tunnel of Corti
\cite{Karavitaki2007} or on the tectoral membrane
\cite{Ghaffari2007}.

The charge $q$ inside the OHC is controlled by the mechanically
sensitive ion channels of the hair bundle
\cite{Hudspeth2008}. Equation \eqref{eq:charge} models the capacitor
of the OHC that integrates ion currents flowing through the ion
channels. The OHC has an active mechanism to regain its resting
potential, which is modeled by the rate constant $\gamma$. We assume
that the open probability of the ion channels is described by a
second-order Boltzmann function \cite{Lukashkin1998}
\[
  P_\mathrm{O} = \left[ 1 + \e^{-a_1(x-b_1)} \left(1+\e^{-a_2(x-b_2)}\right) \right]^{-1}.
\]
The piezoelectric property of the OHC is represented by the constant $\nu$.

The boundary condition of the model at the apex is determined by the
helicotrema, which we assume has no resistance to fluid motion, hence
$p(1,t)=0$. The boundary condition at the base is determined by
the action of the stapes on the oval window. The stapes is in turn
controlled by the dynamics of the middle ear
which we model as a single-degree-of-freedom oscillator
\begin{equation}
  \ddot{x}_\mathrm{ow} + 2\zeta_\mathrm{ow} \dot{x}_\mathrm{ow} + \omega_\mathrm{ow}^2 x_\mathrm{ow} = m_\mathrm{ow}^{-1}\left( p(0,t) + G p_e \right),
  \label{eq:midear}
\end{equation}
where $p_e$ is the sound pressure acting on the eardrum and the
subscript `$ow$' refers to the oval window. Hence the basal
boundary condition can be determined from force balance to be
\[
  \pdiff{p}{\xi}(0,t) = m k_\mathrm{ow} \ddot{x}_\mathrm{ow}(t).
\]

\subsection{Parameter fitting}

%%%%%% TABLE %%%%%%
\begin{table*}
\caption{Parameter values of the mathematical model}
\begin{tabular*}{\hsize}{rrl}
Parameter & Value & Description \\
\hline \\
$\varepsilon$ & 0.0594 & cochlear geometry constant \\%\cr
$m$ & \unit{0.0055}{\gram\usk\centi\metre\rpsquared} & mass density of the Basilar membrane \\%\cr
$\omega_0$ & $20.832 \mathrm{e}^{-4.8354\xi} - 0.1455$ & undamped natural frequency of the Basilar membrane \\%\cr
$\zeta$ & $0.11\mathrm{e}^{1.4\xi}$ & relative damping of the Basilar membrane\\
$f_q$ & 0.2221 & gain of the outer hair cell \\
$L$ & $0.005 \mathrm{e}^{0.7\xi}$ & characteristic feed-forward distance\\
$\gamma$ & $12.5 \times 5^{-\xi}$ \reciprocal{\milli\second}& RC time constant of the outer hair cell \\
$\nu$ & \unit{0.72}{\nano\coulomb} & piezoelectric constant of the outer hair cell \\
$I_\mathrm{hb}$ & \unit{0.390323}{\nano\ampere} & maximum transduction current through the hair bundles \\
$a_1$ & 19.9873 & parameter of $P_\mathrm{O}$ \\
$a_2$ & 16.3928 & parameter of $P_\mathrm{O}$ \\
$b_1$ & 0.0692021 & parameter of $P_\mathrm{O}$ \\
$b_2$ & 0.0369987 & parameter of $P_\mathrm{O}$ \\
$m_\mathrm{ow}$ & \unit{1.85}{\gram\usk\centi\metre\rpsquared} & surface density of the stapes \\
$k_\mathrm{ow}$ & \unit{1100}{\reciprocal{\centi\metre}}  & coupling of oval window to perilymph \\
$\zeta_\mathrm{ow}$ & 0.265 & relative damping of the middle ear\\
$\beta$ & 0.78 & convolution kernel Gaussian characteristic width \\
$\delta_g$ & 0.74 & convolution kernel Gaussian shift \\
$\delta_o$ & $-$0.04 & convolution kernel wave shift \\
$L_0$ & 0.005 & feed-forward distance at base \\
$\mu_L$ & 0.7 & rate of growth of the feed-forward distance \\
\hline \\%\cr
\end{tabular*}
\label{tab:params}
\end{table*}
\twocolumngrid
%%%%%% TABLE %%%%%%
The model \eqref{eq:fluid}-\eqref{eq:midear} is only complete once
parameters have been specified; a full list is given in Table \ref{tab:params}.
We fitted the model to the frequency response function (FRF) of the BM
at \unit{0}{\deci\bel} sound pressure level (SPL) of animal N92 in the
data of Rhode \cite{Rhode2007}. The sharpness of tuning was fitting simple
exponential functions to $\zeta(\xi)$,
$L(\xi)$ and $\gamma(\xi)$ and by finding a suitable value for
$f_q$.  The shape of the FRF was tuned
by varying the parameters of $w(h)$, that is $\beta$, $\delta_g$ and
$\delta_o$. However, tuning the model at a single position is not
sufficient, because we want the sharpness of tuning to be accurate at
every BM position. Unfortunately this type
of data from a single cochlea does not currently exist.
Reasoning that BM tuning and neural
tuning are close, we instead use the neural tuning data
of Recio-Spinoso {\em et al.}\ \cite{RecioSpinoso2005}
to further fit the parameters $\zeta(\xi)$,
$L(\xi)$ and $\gamma(\xi)$ of our model.

The nonlinear response of the model was tuned at the CF position of
\unit{6.6}{\kilo\hertz} for the \unit{6.6}{\kilo\hertz} stimulus. Only
the parameters $a_1$, $b_1$, $a_2$ and $b_2$ of the open probability
function $P_\mathrm{O}$ were altered, while keeping the derivative at
the equilibrium $\dash{P}_\mathrm{O}(0)$ constant, so that the
\unit{0}{\deci\bel} FRF persisted.  The sensitivity of the model is
then tuned by adjusting only two parameters: $G$, which scales the
input signal, and $\Delta$, which tunes the sensitivity of the hair
bundles, and effectively determines the BM amplitude range.

\section{Results \& Discussion}

\subsection{BM response}

\begin{figure*}
\begin{center}
\includegraphics[width=0.99\textwidth]{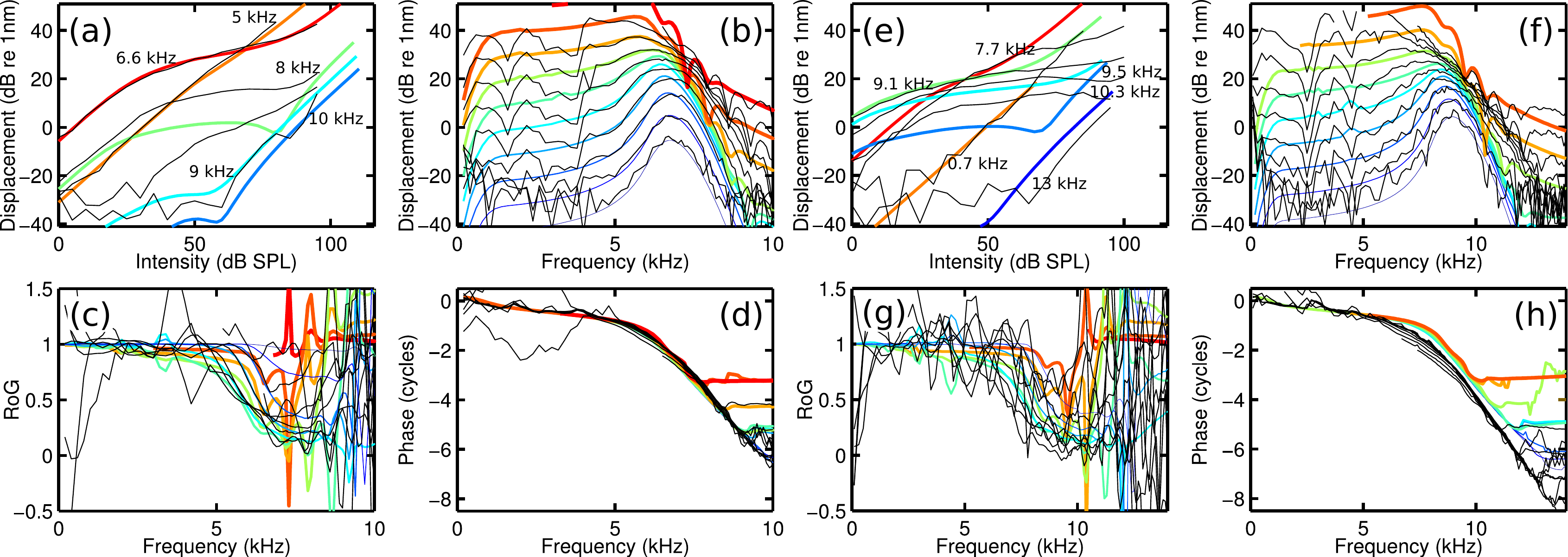}
\end{center}
\caption{BM response of the model compared to data sets in \cite{Rhode2007} taken at the (a-d) CF \unit{6.6}{\kilo\hertz} position and (e-h) at the position with CF \unit{9.1}{\kilo\hertz}. (a,e) BM displacement as a function of sound pressure level at different stimulus frequencies. (b,f) Frequency response functions at different SPLs, the lowest curve corresponds to 0 dB SPL. (c,g) Rate of growth indicating compression at different amplitudes as a function of frequency. (d,h) phase of the BM displacement. Thin black lines represent experimental data published in Rhode \cite{Rhode2007} and thick color lines represent numerical results. The bottom curve in (b-d) and (f-h) corresponds to 0 dB
SPL, while consecutive curves (reading up) increase by 10 dB in SPL.\label{fig:bmresp}}
\end{figure*}

In order to find the BM response, we solve our model
\eqref{eq:fluid}-\eqref{eq:midear} numerically; see the Materials and
Methods section for details.  The results are presented in
Fig.~\ref{fig:bmresp} in the form of thick colored lines. The
superimposed thin black curves show the corresponding experimental
data.
%%%%%%%% HERE FIGURE %%%%%%%%
%%%%%%%% HERE FIGURE %%%%%%%%

Results at the CF place of \unit{6.6}{\kilo\hertz}, for which our
model is tuned, are shown in Figure \ref{fig:bmresp}(a-d).  Figures
\ref{fig:bmresp}(a) and (b) show input--output functions; BM
displacement as a function of sound pressure level (SPL) at various
frequencies, and as a function of SPL at various frequencies,
respectively. The data at lower than CF frequencies are accurately
matched by the results, at higher than CF frequencies the calculated
response is more compressed at lower SPLs than the data. The maximal
compression rate, however, matches the data well. Figure
\ref{fig:bmresp}(b) shows the frequency response functions at
different SPLs. Again, the agreement with data is good. However, for
higher amplitude stimuli the peak in the data widens almost
symmetrically, while our calculations show almost the same width; this
is the same disagreement as in Figure \ref{fig:bmresp}(a).  One of the
reasons why the peak of the tuning curve in our calculations does not
broaden is because our convolution kernal $w$ does not depend on
SPL.  One way to
improve the model would be to allow the parameters of
$w$ to vary with $p_e$. Possible physical causes of such
variation include tectoral membrane waves \cite{Ghaffari2007} and
tunnel of Corti flow \cite{Karavitaki2007}.

Sound compression can be measured by the rate of growth (RoG), defined as the
slope of the log-log graph of BM displacement versus SPL. Figure
\ref{fig:bmresp}(c) shows the RoG, as a function of frequency, for a
range of different stimulus intensities. Our model shows good
agreement with the data. As expected, compression is minimal for
frequencies significantly lower than CF, then increases (so that the
RoG can be as low as $0.1$). Note that, due to interference, the RoG
fluctuates significantly at certain frequencies especially above CF,
which the model also reproduces.  The phase variation of the BM
response is illustrated in Figs.\ \ref{fig:bmresp}(d). Again we see
good agreement between model predictions and experimental data.

Even though our model has been fitted to data from a specific animal
whose BM vibration was measured at the \unit{6.6}{\kilo\hertz} CF
position, we can compare its predictions to another dataset at a
different frequency, measured in a different animal of the same
species. Such data are shown in Figure \ref{fig:bmresp}(e-h).
The model is adjusted
by changing only two parameters: the input sensitivity $G$ and hair
bundle sensitivity $\Delta$. The numerical results are still
reasonably close to the data and all the qualitative features are
preserved. The most obvious differences are that our model shows
slightly sharper tuning than the animal, and that the phase variation
along the frequency axis is somewhat greater in the experimental data
(again possibly a result of the assumption that $w$ is independent of
stimulus).

\subsection{Stability of ideal and rough cochlea models}
Having shown the validity of our model, we now use it to investigate
the stability of the cochlea.
Dynamical systems theory states that if the spectrum of
the system at an equilibrium state has only negative real parts, the system
is stable and small perturbations to that equilibrium will decay. Since
our cochlea model \eqref{eq:fluid}-\eqref{eq:midear} is a partial
differential equation with spatial feedback and feed-forward, its
spectrum will typically contain a continuous curve in the complex
plane, spanning frequencies
from the lowest to the highest audible frequencies.
However, our use of a numerical discretization scheme (which
approximates the system by a large set of ordinary differential
equations) means that we can in practice compute the spectrum using a
standard eigenvalue solver, resulting in a discrete approximation to
the spectrum, with a large number of discrete eigenvalues.

%%%%%%%% HERE FIGURE %%%%%%%%
\begin{figure}
\begin{center}
\includegraphics[width=0.49\textwidth]{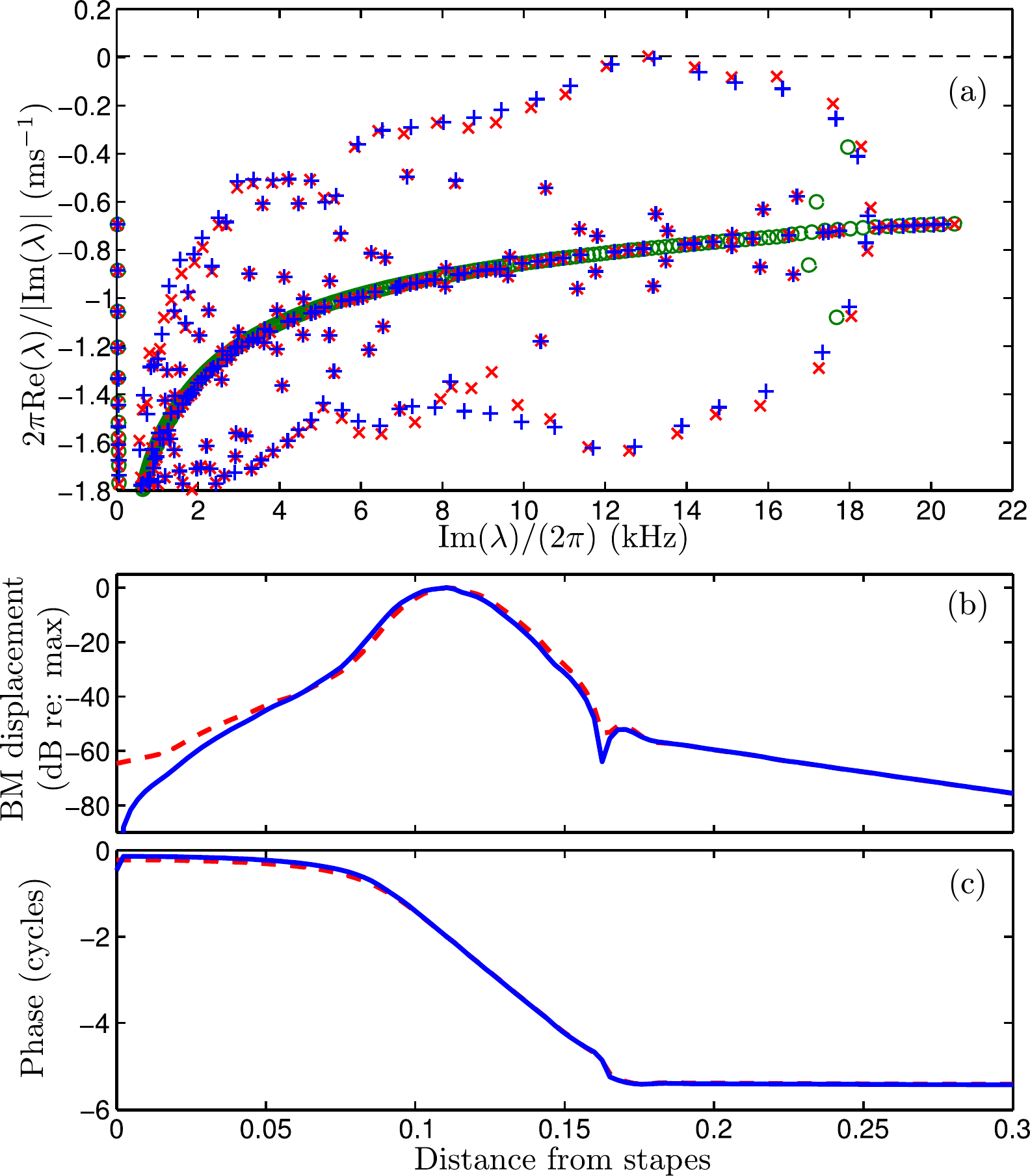}
\end{center}
\caption{(a) Characteristic roots of the cochlear model. Green circles correspond to the smooth cochlea, red crosses to the rough cochlea with $\sigma=0.06$, and blue pluses represent a cochlea of the same roughness without the middle ear.
The horizontal axis is the frequency of the characteristic root and the vertical axis is the relative damping,
which is the log of the amplitude ratio of two consecutive periods of free vibration with the given frequency.
(b,c) Magnitude and phase of the eigenvector corresponding to the most unstable eigenvalue, for the rough cochlea model; red dashed and solid blue lines correspond to models with and without the middle ear respectively. \label{fig:spectrum}}
\end{figure}
%%%%%%%% HERE FIGURE %%%%%%%%
Figure \ref{fig:spectrum}(a) shows computed spectra of the cochlea model, in the absence of stimulus. The imaginary axis is shown as a dashed line;
any eigenvalue $\lambda$ above this line indicates
instability of the cochlea.
The green circles in Figure \ref{fig:spectrum}(a) (mostly organized in
the thick line in the middle of the graph) represent the spectrum of
the cochlea described above; they show that $\Im(\lambda)<0$ for all
$\lambda$, so the cochlea is stable to small disturbances.

This calculation is for an ideal cochlea;
a real organ will in general have small geometric variations of
its properties along its length.
In order to model such inhomogeneities, we allow the feed-forward
parameter $f_q$
to vary randomly about its notional value at each position of the
cochlea. We assume this variation to be
normally distributed with zero mean and constant variance $\sigma^2$.
Such randomness has been shown by Ku et
al.\ \cite{Elliott2009} to induce reflections from the Basilar
membrane back towards the stapes.  In what follows we shall consider
only one realization of the distribution, but we shall allow the
variation to be scaled, in effect varying the standard deviation
$\sigma$. We shall refer to such a model the {\em rough cochlea},
with $\sigma$ a measure of the degree of roughness.

The spectrum of one particular rough cochlea (with $\sigma=0.06$) is
represented in Figure \ref{fig:spectrum}(a) by red $\times$ signs. We
see that the continuous spectrum breaks up; the eigenvalues are
scattered, and discrete eigenvalues jump out of the curve found for
the smooth cochlea, sometimes by a significant distance. The closer an
eigenvalue gets to the imaginary axis, the more the cochlea becomes
sensitive to a disturbance at the corresponding frequency, which
lowers the hearing threshold at that frequency. If an eigenvalue
crosses the imaginary axis, a Hopf bifurcation occurs; as a result the
cochlea becomes unstable and we would expect a spontaneous oscillation
in the absence of any stimulus. This explains why SOAEs occur at the
frequencies where the hearing threshold has notches. We also see in
Figure \ref{fig:spectrum}(a) that spectral points close to the
imaginary axis are roughly periodically spaced in frequency. This
agrees with data that spontaneous and stimulus frequency OAEs and the
hearing threshold microstructures are roughly periodic in frequency
\cite{Kemp1981}.

\subsection{Coherent reflections}
The observation from Figure \ref{fig:spectrum}(a) that spectral points
close to the imaginary axis are regularly spaced in frequency might be
explained by the theory of Zweig and Shera \cite{ZweigShera1995},
whose assumption is that resonances build up in the cochlea between
the oval window and the CF position of the resonance. To test such a
hypothesis we
removed the middle ear from our model altogether, and made sure that
there is no reflection from the oval window by setting
$p(0,t)=0$. Calculating the stability of this model, with the same
roughness as before, yields a remarkably similar spectrum (illustrated
by the blue $+$ signs in Figure \ref{fig:spectrum}(a)). This means
that, while reflections from the oval window play a part in shaping
resonances, they are not essential.

To further clarify how reflections occur we also calculated the
eigenvector that corresponds to the largest
$\mathrm{Re}(\lambda)/\vert \mathrm{Im}(\lambda) \vert$ value; shown
in Figure \ref{fig:spectrum}(b,c), with and without the middle ear
(red dashed and blue solid curves respectively). These correspond to
the vibration pattern of the most unstable mode of the cochlea. It can
be seen that the magnitude of the vibration rapidly decreases towards
the base of the cochlea when the middle ear is removed. However, the
BM pattern is remarkably similar to that with the intact middle ear
for the rest of the cochlea.

To explore the implications of local reflections, we also considered
the case when the longitudinal coupling parameter $f_q$ is increased
by 20\% at a single position, so that it forms a step function
from base to apex, rather than varying randomly with length (data not shown).
We might imagine that such an inhomogeneity will act as a single point
of reflection. We found instead that distinct resonances were
found (confirming the result in Ku {\em et al.}\ \cite{Elliott2009}), and the
cochlea patterns at the strongest resonance were indistinguishable
from the result in Figure \ref{fig:spectrum}(b,c).
We therefore conclude that the theory of coherent
reflections is not adequate to explain our results.

\subsection{Spontaneous emissions}

%%%%%%%% HERE FIGURE %%%%%%%%
\begin{figure}
\begin{center}
\includegraphics[width=0.49\textwidth]{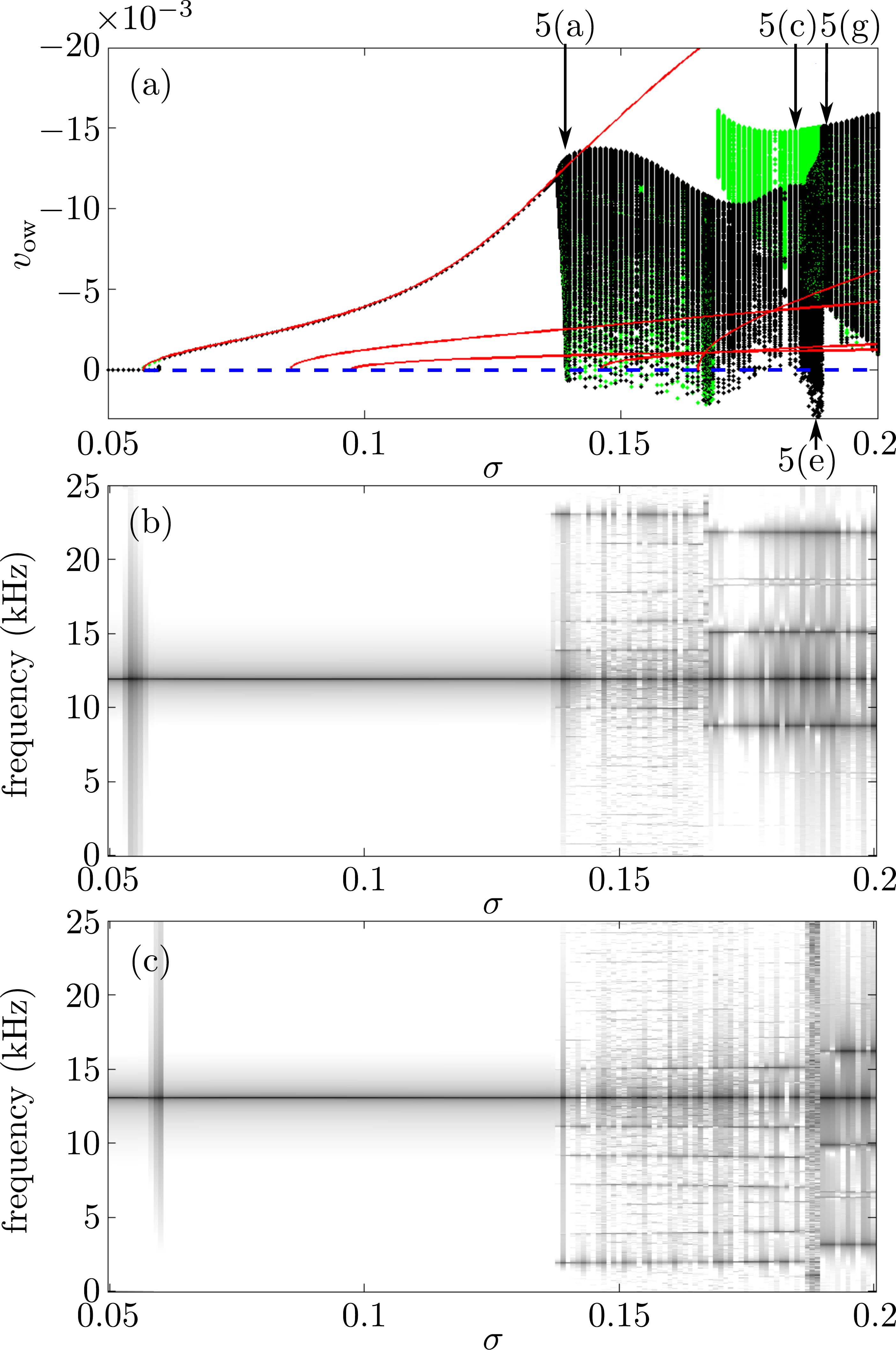}
\end{center}
\caption{(a) Amplitude of spontaneous emissions calculated from the model,
represented by oval window fluid velocity as a function of surface roughness.
Red (periodic orbits) and blue (fixed point) curves denote the results of numerical continuation of the model. Black and green dots represent simulation results when increasing and decreasing $\sigma$, respectively. Points indicated by
arrows represent responses plotted in detail in Figure \ref{fig:simres}.
(b,c) Spectrograms of the cochlea emission with decreasing $\sigma$  and increasing $\sigma$ respectively.\label{fig:bifdiag}}
\end{figure}
%%%%%%%% HERE FIGURE %%%%%%%%
We can also use our unforced cochlear model to quantify the nature of
the spontaneous oscillations that exist beyond any point of
instability.  When an eigenvalue crosses the imaginary axis, as in the
case of the rough unforced cochlea model described above, a Hopf
bifurcation (of the entire cochlea) occurs; as a result, we expect a
spontaneous oscillation to develop at approximately the frequency of the
eigenvalue. Figure \ref{fig:bifdiag}(a) shows the results of
a combination of numerical continuation and
simulation techniques to track both stable and unstable periodic
orbits, and also to reveal more complicated dynamical behavior.
The resting position of the cochlea is represented by a blue line, the
solid portion representing where this is stable and the dashed portion
where it is unstable.
The red curves show the results of tracking the limit cycle motions
born at a sequence of Hopf
bifurcations that occur as the surface roughness $\sigma$ increases.
shows a graph of the cochlear response as $\sigma$ varies.
The vertical axis shows the stapes velocity $v_{ow}=\dot{x}_{ow}$ at the
time instance when the stapes displacement is zero.
This creates a
sequence of values $v_{ow}$ which can be represented as a
so-called Poincar\'{e} map $v_{ow}\mapsto P(v_{ow})$. Superimposed on the
plot are the results of direct simulations, after any transient motion
has decayed. These are represented by black and green dots. The black dots
were obtained by making a slow forward sweep in $\sigma$ and the green
dots by making a backward sweep.
Corresponding spectrograms are shown in Figure \ref{fig:bifdiag}(b,c).

Note from Figure \ref{fig:bifdiag}(a) that as $\sigma$ increases from
zero, the unforced cochlea goes unstable at around $\sigma=0.05$, via a
supercritical Hopf bifurcation.  Thus, a stable periodic orbit is born
as $\sigma$ increases through this value. This represents a single
frequency SOAE at $\approx$\unit{13.072}{\kilo\hertz},
seen clearly in the spectrograms.
Further supercritical Hopf bifurcations occur from the
(now unstable) steady state as $\sigma$ increases further, although
these branches of limit cycles all appear to be unstable.
The single stable periodic solution can be inferred from where
the where black dots of the simulation data overlie the
red line from numerical continuation.
The amplitude of this limit cycle grows steadily with $\sigma$.

Suddenly, for $\sigma \approx 0.13738$, the (single frequency) periodic orbit
loses stability at $\sigma = 0.13738$, through a supercritical secondary Hopf
(or Neimark-Sacker) bifurcation. This creates quasi-periodic motion
represented by the wide region of black dots in Figure \ref{fig:bifdiag}(a).
A second frequency \unit{1.93}{\kilo\hertz} is clearly
visible in the spectrograms.
At $\sigma \approx 0.168$ a second quasi-periodic orbit appears,
with frequencies 3.16 and \unit{13.06}{\kilo\hertz}, which coexists with the first for larger values of $\sigma$. The two different attractors are identified by different patterns of black and green dots in Figure \ref{fig:bifdiag}(a) for $0.168<\sigma<0.186$, and different patterns in the spectrograms in Figure \ref{fig:bifdiag}(b,c).
At $\sigma \approx 0.186$ the first quasi-periodic orbit (black dots) breaks
up to form a chaotic attractor; this is indicated by a
sudden broadening of the spectrogram in Figure \ref{fig:bifdiag}(c).

%%%%%%%% HERE FIGURE %%%%%%%%
\begin{figure*}
\begin{center}
\includegraphics[width=\textwidth]{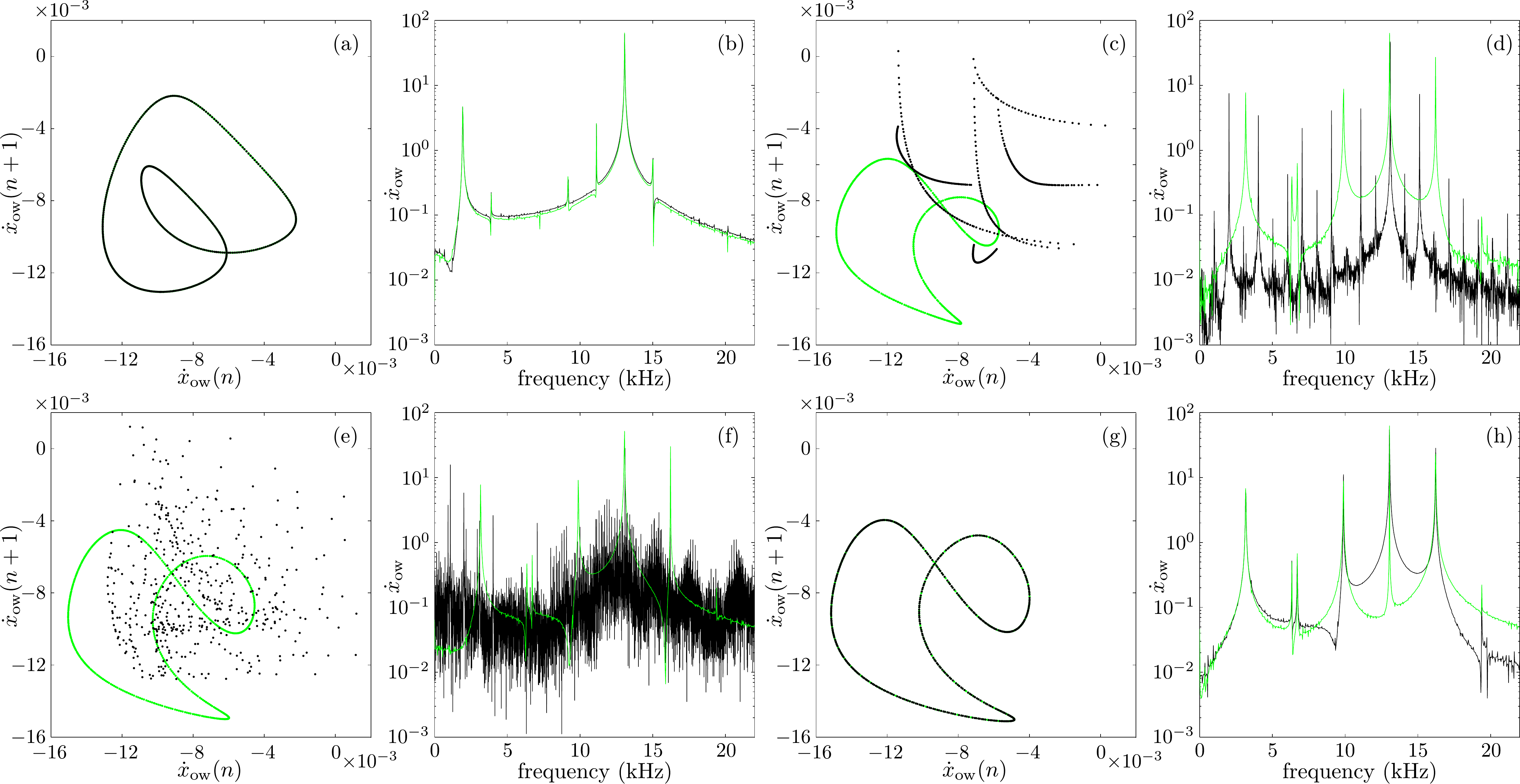}
\end{center}
\caption{Poincare maps (a,c,e,g) and corresponding frequency spectra (b,d,f,h) of solutions of the cochlear model, for values of cochlear roughness $\sigma$ indicated in Figure \ref{fig:bifdiag}(a).
See text for details. \label{fig:simres}}
\end{figure*}
%%%%%%%% HERE FIGURE %%%%%%%%
Figure \ref{fig:simres} illustrates the qualitatively different
classes of emissions predicted by the model in more detail, showing
both the orbits (as Poincar\'{e} maps) and their spectra.  The first
quasi-periodic orbit, just after onset, is shown in Figure
\ref{fig:simres}(a-b) and the two coexisting quasi-periodic orbits by
green and black dots in Figure \ref{fig:simres}(c-d); note how the
differing second frequencies result in different spacings between the
peaks in the spectra. The chaotic attractor, shown with black dots in
Figure \ref{fig:simres}(e-f) is of broadband character. Finally, the
chaotic attractor disappears at $\sigma=0.1892$, where only a single
quasi-periodic orbit remains, shown in Figure \ref{fig:simres}(g-h).

The scenario described above can change significantly when applying
different roughness patterns to the feedback coefficient
$f_q$. However, simulations with different realizations (not shown)
suggest that the supercritical nature
of the bifurcation remains, and that supercriticality is the property
of the chosen open probability function $P_\mathrm{O}$. Changing
$P_\mathrm{O}$ can result in a subcritical Hopf bifurcation, as
reported in \cite{SzalaiMoH2011}.

The result that at a single parameter value
multiple stable vibration patterns
can coexist has implications for explaining other
features observed experimentally. A large enough perturbation (e.g., a
click) can push the ear into exhibiting different SOAE spectra, making
one SOAE frequency appear and another disappear. Such changes are
found experimentally \cite{Burns2009}, as are hallmarks of chaotic
oscillations \cite{Keefe1990}, in qualitative agreement with the
dynamics we predict.

\section{Conclusions}
In this paper we have introduced a nonlinear transmission-line model
of the mammalian hearing organ.  We believe this to be the first
mathematical model that is both able to capture the turning curves of
the Basilar membrane across a range of different frequencies and come
up with credible, testable explanations for observed finite-amplitude
otoacoustic emissions.  Specifically, the model provides a
parsimonious synthesis of many of the features that have been proved
important in previous modeling and experimental studies over several
decades: fluid-structure interaction leading to waves that travel a
frequency dependent distance to the characteristic frequency position
along the Basilar membrane; longitudinal variation in organ of Corti
properties; local nonlinearity due to the combined effects of outer hair
cell somatic motility and hair bundle adaptation; and spatial
feed-forward effects due to longitudinal coupling within the OOC.

The results in Figure \ref{fig:bmresp} show how the combination of
longitudinal coupling together with local nonlinearity is able to
reproduce both qualitatively and quantitatively BM vibration data and
neural tuning in different animals. We are not aware that models with
either pure longitudinal coupling or pure local nonlinearity are able
to reproduce such features.

Moreover, by applying a range of techniques from dynamical systems
theory, we have been able to calculate both instability thresholds and
the waveforms of post-instability spontaneous otoacoustic emissions. We
have shown how the roughness of the cochlea can be responsible for
producing these resonant instabilities.  Moreover, by removing the
middle ear and allowing no reflections at the oval window, we found
that the resonance to only slightly diminish, but for the overall
pattern of BM motion to be maintained. This shows that SOAEs can be
produced by instabilities that do not require reflections from the
middle ear. Reflections, however, might occur elsewhere.

Our model also how spontaneous OAEs arise robustly from supercritical
Hopf bifurcations of the entire cochlea.  Thus we do not need to make
artificial ``Hopf ear'' hypotheses about local nonlinearities at a
particular CF being responsible.  Furthermore, with significant
roughness, the model exhibits a range of complex emissions beyond
pure tones including chaotic signals and bi-stability between different
kinds of SOAE. These properties can explain reported recordings of
emissions that appear to spontaneously switch between different signal
patterns.

\section{materials}
\subsection{Equivalent formulation of the model}
In order to be able to discretize our model into a set of ordinary differential equations, we first combine \eqref{eq:fluid} and \eqref{eq:BM} to eliminate the pressure $p$
\begin{multline}
  \ddot{x}_\mathrm{bm} - \varepsilon^2 \frac{\partial^{2}}{\partial\xi^{2}} \ddot{x}_\mathrm{bm} = \left. \varepsilon^2 \frac{\partial^{2}}{\partial\xi^{2}} \right(2\zeta\omega_{0}\dot{x}_\mathrm{bm}(\xi,t) + \omega_{0}^{2} x_\mathrm{bm}(\xi,t)  \\
  \left. + \frac{f_{q} \omega_{0}^{2}}{L} \int w(h / L) q \left(\xi - h ,t\right)\, \mathrm{d}h
  \right).\label{eq:rewrit}
\end{multline}
Equation \eqref{eq:charge} stays unmodified. The governing equations of the middle ear are rewritten in state-space form using $x_1 = x_\mathrm{bm}(0,t)$, $x_2 = \dot{x}_\mathrm{bm}(0,t)$, $y_1 = x_\mathrm{ow}$ and $y_3 = \dot{x}_\mathrm{ow} - (m/m_\mathrm{ow})\dot{x}_\mathrm{bm}(0,t)$ as follows
\begin{align*}
\dot{y}_{1}&=y_{3}+\kappa^{-1}x_{2},\nonumber \\
\dot{y}_{3}&=-\omega_\mathrm{ow}^{2}y_{1}-2\zeta_\mathrm{ow}\omega_\mathrm{ow}y_{3}+\kappa^{-1}\omega_\mathrm{bm}^{2}x_{1}+G_{e}p_{e},\nonumber\\
\dot{x}_{1}&=x_{2},\\
\dot{x}_{2}-\delta\frac{\partial}{\partial x}\dot{x}_2 &=\delta\frac{\partial}{\partial x}\left(2\zeta_\mathrm{bm}\omega_\mathrm{bm}x_{2}+\omega_\mathrm{bm}^{2}x_{1}\right)\nonumber\\
   &\quad -\left(2\zeta_\mathrm{bm}\omega_\mathrm{bm}-2\zeta_\mathrm{ow}\omega_\mathrm{ow}\right)x_{2}-\omega_\mathrm{bm}^{2}x_{1}\nonumber\\
   &\quad -\kappa\left(G_{e}p_{e}-2\zeta_\mathrm{ow}\omega_\mathrm{ow}y_{3}-\omega_\mathrm{ow}^{2}y_{1}\right),\nonumber
\end{align*}
where $\kappa=m_\mathrm{ow}/m$, $\delta=\kappa/k_\mathrm{ow}$ and $G_e=G/m_\mathrm{ow}$.

\subsection{Numerical methods}
We discretize our equations along the length of the cochlea using
finite differences. We use a non-uniform mesh that has interval length
between the mesh points proportional to $L(\xi)$. The first space
derivative is a backward looking finite difference,
and the second derivative is obtained using central differencing.
The integral representing the feed-forward is approximated by the
trapezoid rule. Discretizing the governing equations with this scheme
yields a set of ordinary differential equations that are solved using
the MATLAB routine \texttt{ode113}, specifying a constant mass matrix
that arises from the second spatial derivatives of the right hand side
of \eqref{eq:rewrit}. The relative error tolerance was set to
$10^{-10}$, and the absolute tolerance to $10^{-11}$.

The numerical continuation to obtain both the pure-tone response in
Figure \ref{fig:bifdiag} and the periodic orbits branching from the
Hopf bifurcation points are calculated from a periodic boundary value
problem in time. The temporal discretization is performed using
orthogonal collocation \cite{deBoor1973} with 4th degree interpolating
polynomials on 12 equidistantly spaced intervals. We used pseudo
arclength continuation \cite{Doedel1991} to detect bifurcation points,
and grow branches of solutions.

\end{document}